# Distribution of Transport Current in a Type II Superconductor Studied by Small Angle Neutron Scattering.

A. Pautrat, C. Goupil, Ch. Simon
*Laboratoire CRISMAT, UMR 6508 du CNRS et de l'Institut
Supérieur de la Matière et du Rayonnement, 14050 CAEN, France.*

D. Charalambous, E.M. Forgan
*School of Physics and Astronomy, University of Birmingham,
Birmingham B15 2TT, United Kingdom.*

G. Lazard, P. Mathieu
*Laboratoire de Physique de la Matière Condensée de l'Ecole Normale Supérieure,
UMR 8551 du CNRS, associée aux universités Paris 6 et 7,
24 rue Lhomond, F-75231 Paris Cedex5, France.*

A. Brûlet
*Laboratoire Léon Brilloin,
CEN Saclay, 91191 Gif/Yvette, France.*

(Dated: February 20, 2003)

We report Small-Angle Neutron Scattering (SANS) measurements on the vortex lattice in a PbIn polycrystal in the presence of an applied current. Using the rocking curves as a probe of the distribution of current in the sample, we observe that vortex pinning is due to the surface roughness. This leads to a surface current that persists in the flux flow region. We show the influence of surface treatments on the distribution of this current.

The nature of the pinning of the Flux Line Lattice (FLL) in superconductors is still the subject of intense discussion. In most of the literature, it is assumed that vortex pinning in clean enough superconductors is due to small defects in the bulk (collective bulk pinning)[1, 2]. On the other hand, the dominant role of surface roughness has also been emphasised ([3, 4, 6] and older references therein). Perhaps one of the reasons for this old and unresolved debate arises from the difficulty in finding discriminating experiments, which can unambiguously separate bulk and surface currents. As an example, the classical Bean method—relying on the magnetization hysteresis curve—postulates, but does not demonstrate, the existence of a bulk critical current density. It is also possible to check in a pickup coil the distribution of a current modulated at low frequency. However, such measurements are not conclusive if one uses too low a current since, even in the case of bulk pinning, a small ac excitation is screened by the well-known Campbell penetration depth [5, 6]. It is necessary to measure the complete frequency dependence of the penetration length, in a sample of well-defined shape, in order to have good test of the predictions [6]. A further problem is the difficulty of treating the linear ac response of the vortex lattice [6, 7]. In this paper, we use another technique, based on neutron diffraction by the vortex lattice [8, 9]. This provides direct information about the flux line structures and shapes in the bulk of the sample, in the presence of a large dc current. As a consequence of the Maxwell-Ampère equation:

$$\nabla \times \boldsymbol{B} = \mu_0 \boldsymbol{J}, \qquad (1)$$

The penetration of the current in the bulk of the sample leads to a curvature of the field lines. This can be measured by neutron diffraction, as a broadening of the relevant rocking curve of a Bragg peak, as observed a long time ago by Schelten et al. [10]. As we will describe below, it is then possible to discriminate between surface and bulk critical current effects.

The present experiments were performed on two polycrystalline PbIn samples, both containing 10.5% of In by weight ($T_c = 6.9$ K). Their detailed preparation can be found elsewhere [11]. After annealing to obtain homogeneous bulk properties, the samples were pressed between glass plates. The resulting surface polish at the optical scale reduces the critical current, as previously reported [12]. We note that our sample is polycrystalline with grains of about $100\,\mu$m in size. It is therefore not free from bulk defects. It is usually assumed that surface pinning will be important only in superconductors without bulk defects and with a particularly rough surface. We find experimentally that this assumption is not justified. The dimensions of the two PbIn samples were respectively $\ell \times w \times t = 30 \times 5.5 \times 0.5 \,\text{mm}^3$ and $30 \times 6 \times 0.6 \,\text{mm}^3$. Surface treatments were performed on the samples in order to modify their critical currents. After a reversible increase of the surface roughness by mechanical "abrading" by gently wiping the surfaces with a commercial solution containing micrometers particles, the polish at the optical scale disappears and a huge increase of critical current is observed (up to a factor of ten). The first sample was measured by neutron diffraction before and after this treatment. Using a mixture of hydrogen peroxide and acetic acid as a chemical polish, we could recover the mirror aspect of the surface and decrease the critical current again. Solely by surface treatment, it is also possible to obtain critical current values which vary strongly on the scale of the sample dimensions. This was the case for the second sample. One should also point out that in the case of a homogeneous critical current, the voltage-current ($V(I)$) characteristics are straight lines, shifted from the origin by an amount equal to the critical current. An explanation of such a linear $V(I)$ curve ($\frac{dV}{dI}$ =constant (ct)) is not as trivial as it would seem. It exhibits a permanent critical current even at high vortex velocity and hence is by definition non-ohmic ($\frac{V}{I} \neq$ ct). The consistency of a linear $V(I)$ curve with a surface pinning model has been demonstrated [3], but an explanation using bulk pinning models and elastic instabilities has not been conclusive [13, 14]. Moreover, recent theoretical treatments have shown that the effect of the pinning potential should decrease rapidly with the vortex lattice velocity and tend to an ohmic-like response ($\frac{V}{I} \approx$ ct) [15, 16], in contradiction with our experimental curves. The central question is how does the vortex lattice keep a memory of the pinning potential when it is ordered and free to move? The theoretical answer depends on the way that the critical part and the dissipative part of the current are treated [3], and it is of particular importance when writing dynamical equations. The present experiment provides evidence that a clear separation between these two currents can be made.

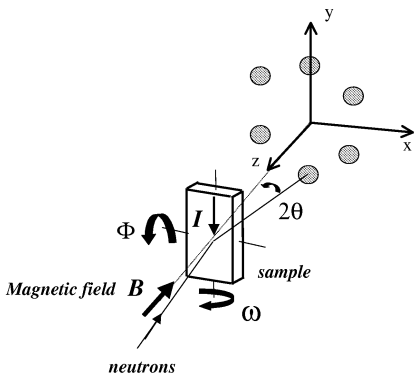

FIG. 1: The geometry of the SANS experiment used to observe diffraction by the vortex lattice in the presence of an applied current. To obtain the rocking curves described in the text, sample and cryomagnet are rocked together by angles $\phi$ or $\omega$.

The SANS measurements were carried out using the PAXY instrument at the LLB (Saclay, France). The sample edges were masked with Cd foil to expose an area of $\sim 2 \times 1\,\mathrm{cm}^2$. This removed small-angle scattering from electrical contacts and any signal from distorted FL structures near the edges of the sample. The neutron wavelength used was $\lambda_n = 10$ or $15\,\text{Å}$ (with $\Delta\lambda_n/\lambda_n = 10\%$) with an incident beam divergence of $0.15°$. The magnetic field and neutron beam were almost parallel and the transport current was applied vertically along the length of the sample (see Fig. 1). Thus the Lorentz force acting on the FLL was horizontal, perpendicular both to the magnetic field and transport current. All the experiments were carried out after a field cooling process to avoid any significant field gradients in the sample [17]. The measurements were carried out in superfluid helium at $2.1\,\mathrm{K}$ to reduce sample heating in the flux flow regime. An estimate of the Kapitza resistance between sample and liquid helium gives a temperature rise of less than $0.1\,\mathrm{K}$ for the maximum applied current ($I \simeq 30\,\mathrm{A}$). Information on the FLL structure can be obtained by measuring the "rocking curves" of diffracted intensity versus angle as the sample is rotated about either a horizontal or vertical axis (see Fig. 1). This is of particular importance since the effect of the transport current is strongly anisotropic.

A typical diffraction pattern obtained after field cooling to $T = 2.1\,\mathrm{K}$ at $B = 0.05$, $0.1$ or $0.2\,\mathrm{T}$ ($B_{c2} = 0.48\,\mathrm{T}$) is in the form of a ring. The radius corresponds to the inter-vortex distance of a triangular vortex lattice given by $a_0 = \left((2/\sqrt{3})(\Phi_0/B)\right)^{1/2}$. This "powder pattern" from the FLL is certainly caused by the polycrystalline nature of the Pb-In sample. However, FLL order can be obtained by passing a current larger than critical [18] or by choosing a single crystal of Pb-In [19].

According to equation (1), the presence of current inside the sample has a direct effect on the shape of the field lines. A bulk dc transport current $I$ induces a field $b_x = \mu_0 I/2w$ at the sample surface. The field lines inside the sample are bent with maximum tilt angles given by $\theta = \pm b_x/B$. In our geometry, this results in a broadening of the $\omega$ rocking curve in the lateral Bragg peaks, and for large $\theta$ a good approximation is (see Fig. 2):

$$HWHM(\omega) = \Delta\omega = \theta = \mu_0 I_{bulk}/2wB \qquad (2)$$

The flat top shape of the rocking curve is simply explained by considering a constant curvature of the field lines (and of the corresponding Bragg planes) over the illuminated part of the sample. This is expected if the bulk current is homogeneous. Consistently, and because the curvature of the Bragg planes is in the $x$-direction, the width of the $\phi$ rocking curves ($y$-direction, not shown on the figures) do not change with current, within the error of the fits [18]. Figures 3a and 3b show both the $V(I)$ curves and the broadening of the rocking curves as a function of the current for two different surface states of the first sample. The small critical current ($I_c \approx 1.5$A at 0.2T and $\approx 4$A at 0.1T) corresponds to the smooth surface, and the high critical current ($I_c \approx 12$A at 0.1T) to the abraded and rough surface. Obviously, the surface abrading induces a huge increase of the critical current, but in both cases, the broadening of the rocking curve is given by equation (2), if one replaces $I_{bulk}$ by $(I - I_c)$. This absence of any broadening below $I_c$ gives a direct proof of the absence of any bulk current. One must also emphasize that the bulk current penetrating above $I_c$ is not the total transport current but only $I - I_c$, which corresponds to the bulk dissipative part. This gives a simple explanation for the observed $V = R_{ff}(I - I_c)$ [3], with $R_{ff}$ as a linear flux-flow resistance. All these facts strongly indicate vortex pinning by the surface roughness of the sample. We note that neutron scattering experiments have already given evidence that the vortex lattice structure is not changed by a sub critical current (in Nb [9], in 2H-NbSe$_2$ [20]), in strong contradiction with usual assumptions of the Larkin-Ovchinikov bulk pinning model. The simplest explanation, first suggested by Thorel et al [9], is that the current has not penetrated the bulk of these samples, and hence that there is no reason to look for a bulk pinning force to compensate a bulk Lorentz force.

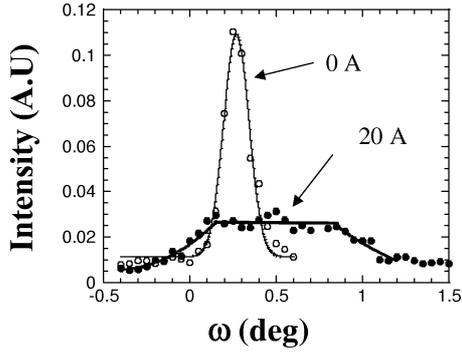

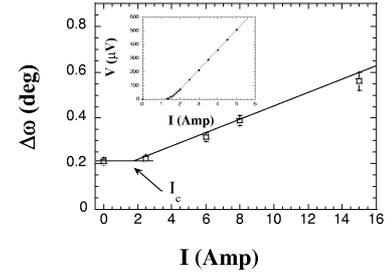

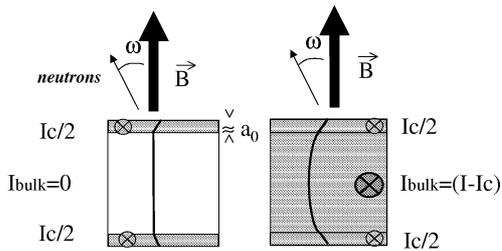

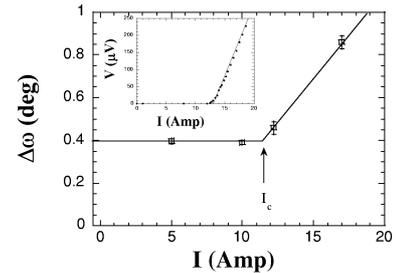

FIG. 2: (a) The $\omega$ rocking curves obtained with and without dc bulk current (first sample with the smooth surface at $B = 0.2T$). They arise from straight field lines (zero bulk current) and from curved field lines (homogeneous bulk current). (b) The distribution of current and the corresponding shape of the field lines for currents above and below the critical value.

FIG. 3: The broadening $\Delta\omega$ of the rocking curves as a function of the applied current for the two different surface states of the first PbIn sample (3a: smooth surface, $B = 0.2$ T and 3b: rough surface and $B = 0.1$T). The slopes of the straight lines for $I > I_c$ are 0.030 and 0.081 deg/A, respectively (compared to 0.033 and 0.065 deg/A calculated with the aid of Eq. 2). Inset: the corresponding $V(I)$ curves.

Fig. 4 shows a $V(I)$ curve for the second sample, which has been treated in order to have an inhomogeneous surface state, with macroscopic rough domains of about $100\mu m^2$. Consequently, the critical current is non-homogeneous on the scale of the length of the sample. This results in a strong curvature of the $V(I)$ characteristic. It was observed a long time ago that this apparent curvature can be expressed as the sum of linear voltage vs current characteristics, all of them possessing a different critical current [21], i.e. the dissipative process could still be linear in nature. This is simply due to the extremely small shear strength of the vortex lattice, which allows the progressive depinning of strips of the vortex lattice with the length of the strips mainly parallel to the vortex velocity. One expects in such a case a bulk penetration of the current in different parts of the sample for different current values. For a given point of the $V(I)$ curve, this leads to a macroscopically non-homogeneous current density in the sample, corresponding to variation in the bending of field lines. As a consequence, the experimental rocking curves no longer have flat top shapes, but present a Gaussian aspect. Let us define a distribution of critical current values $D(I_c)$, normalised so that the total length of the sample is given by $\ell = \int D(I_c)\,dI_c$. Then assuming linear dissipation in each strip, the $V(I)$ curve is given by:

$$V(I) = \int_{I_{c\,min}}^{I} (I - I_c) D(I_c) \frac{R_{ff}}{\ell} dI_c. \qquad (3)$$

Simple differentiation of (3) gives

$$D(I_c) = \left.\frac{d^2 V}{dI^2}\right|_{I_c} \frac{\ell}{R_{ff}}, \qquad (4)$$

which allows $D$ to be obtained from the experimental $V(I)$ curve. Let us note that in a case of a parabolic $V(I)$ curve, which is a good approximation to the non-linear part of the experimental curve (see Fig. 4), $d^2V/dI^2$ is a constant and one finds a constant value of $D \sim 2.10^3\,\mu\mathrm{m}\mathrm{A}^{-1}$. The expected broadening of the rocking curve can be estimated as follow:

$$\Delta\omega = \mu_0 V/(2wBR_{ff}) \qquad (5)$$

As seen in Fig. 4, a direct calculation using this simple model gives excellent agreement.

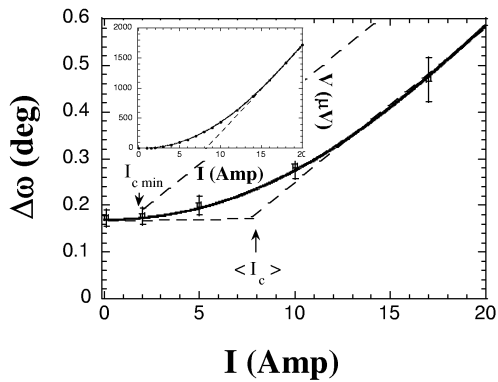

FIG. 4: The broadening $\Delta\omega$ of the rocking curves as a function of the transport current for the Pb-In sample possessing the non-homogeneous surface state ($B = 0.2\mathrm{T}$). The full curve is the prediction of equation (5) taking into account the distribution of surface critical current. The high current linear part gives a slope of 0.032 deg/A compared to the calculated value of 0.030 deg/A. The two dotted lines were obtained using Eq. 2 assuming either a homogeneous critical current of $I_{c,\min} \approx 1.5\mathrm{A}$ (as in Fig. 3a) or of the mean value $\langle I_c \rangle$. The $V(I)$ curve extrapolates to $\langle I_c \rangle$. Inset: the corresponding $V(I)$ curve.

As an illustration, figure 5 shows a simple representation of a $V(I)$ curve in the case of two macroscopic domains possessing different critical currents. This shows that one has to be particularly careful when interpreting such curved $V(I)$ (the rounded onset) in terms of an activated process involving a current dependent depinning energy. As clearly shown here, it could be simply due to variations of the critical current over the length of the sample. In the present experiment, the surface state has been specially treated in order to have a variable critical current. In practice, a perfect homogeneous surface state may be difficult to obtain without special care. Since there frequently exists local variation of purity or of stoichiometry in a sample resulting in small dispersion of the thermodynamic parameters, as in HTc's, the effects of non-uniform critical currents may be more general.

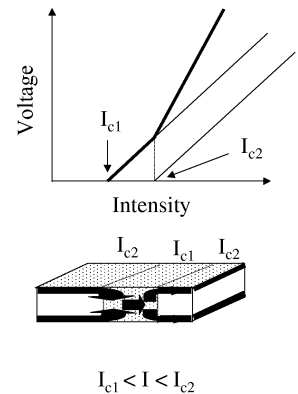

FIG. 5: Schematic representation of a $V(I)$ curve in the case of three domains possessing different critical current values and the corresponding penetration of the current in the bulk of the sample.

## CONCLUSION

In conclusion, by studying the influence of transport current on the rocking curves of the Bragg peaks of the vortex lattice, direct evidence is obtained for surface critical current $I_c$ and bulk dissipative current $(I - I_c)$, and hence surface pinning. A non-linear $V(I)$ curve has been obtained by surface treatment, and is explained by a distribution of surface critical current resulting in non homogeneous bulk current. This experiment sheds light on the behaviour of the transport current and on its separation into a surface critical part $I_c$ and a bulk dissipative part $(I - I_c)$ in a type II superconductor.

Acknowledgments: We would like to thank Frederic Ott (LLB-Saclay) for his help during the experiments.